\documentclass[prd,aps,preprintnumbers,fleqn,showpacs,nofootinbib,superscriptaddress]{revtex4}
\usepackage{amssymb,amsmath,graphicx,epsfig,bm,axodraw}
\usepackage{float}
\usepackage{color} 
\usepackage[caption=false]{subfig} 
\DeclareMathOperator{\tr}{Tr}

\renewcommand{\d}{\mathrm{d}}

\newcommand{\sT}{{\scriptscriptstyle T}}

\newcommand{\pT}{\bm{p}_\sT}

\newcommand{\simorder}{\raisebox{-4pt}{$\, \stackrel{\textstyle >}{\sim} \,$}}

\def\slash#1{\setbox0=\hbox{$#1$}               
        \dimen0=\wd0                            
        \setbox1=\hbox{/} \dimen1=\wd1          
        \ifdim\dimen0>\dimen1                   
        \rlap{\hbox to \dimen0{\hfil/\hfil}}    
        #1                                      
        \else              
        \rlap{\hbox to \dimen1{\hfil$#1$\hfil}} 
        /                                       
        \fi}                                    %

\begin{document}

\preprint{NIKHEF-2014-049}
\title{Impact of gluon polarization on Higgs plus jet production at the LHC}

\author{Dani\"el Boer}
\email{d.boer@rug.nl}
\affiliation{Van Swinderen Institute, University of Groningen, Nijenborgh 4, 9747 AG Groningen, The Netherlands}

\author{Cristian Pisano}
\email{c.pisano@nikhef.nl}
\affiliation{Nikhef and Department of Physics and Astronomy, VU University Amsterdam, De Boelelaan 1081, NL-1081 HV Amsterdam, The Netherlands}

\begin{abstract}
In this paper we consider Higgs plus jet production as a process that is sensitive to the linear polarization of gluons inside the unpolarized protons of the LHC. 
The leading order expressions for the transverse momentum distribution of the Higgs plus jet pair are provided in terms of transverse momentum dependent 
quark and gluon distributions. This includes both angular independent and azimuthal angular dependent contributions, presented directly in the laboratory frame. 
Lacking experimental constraints on the linearly polarized gluon distribution, we study its effects on Higgs plus jet production using two different models to illustrate the generic features and maximal effects. It is found that the $\cos 2 \phi$ distribution may be the most promising observable, as it is driven by only one initial linearly polarized gluon. The potential advantages of the Higgs plus jet process compared to other processes sensitive to the linear polarization of gluons are discussed.    
\end{abstract}

\date{\today}

\maketitle

\section{Introduction}

Higgs production has been shown to be sensitive to the polarization of gluons, even in collisions between unpolarized protons such as 
at the LHC \cite{Catani:2010pd,Sun:2011iw,Boer:2011kf,Boer:2013fca}. 
Gluons with nonzero transverse momentum with respect to the proton momentum can be linearly polarized \cite{Mulders:2000sh}, which affects for instance the transverse momentum distribution of produced Higgs bosons. Although the amount of polarization is currently unknown, it is known that 
it is at the very least perturbatively generated \cite{Nadolsky:2007ba,Catani:2010pd} and, therefore, nonzero. 
There are also strong indications that at small momentum fractions $x$ 
the linear polarization becomes maximal \cite{Metz:2011wb,Dominguez:2011br,Schafer:2012yx}. 
If sufficiently large, the polarization offers a new tool to analyze Higgs couplings to the various Standard Model particles into which it can decay
\cite{Boer:2013fca}. To describe the effects of this gluon polarization, the formalism of transverse momentum dependent (TMD) parton distribution functions (or TMDs, for short) is natural to consider, cf.\ e.g.\ \cite{Collins:2011zzd,GarciaEchevarria:2011rb}. This has been studied for the particular case of Higgs production in Refs.~\cite{Sun:2011iw,Boer:2011kf,Boer:2013fca} and including the effects of TMD evolution in Refs.~\cite{Boer:2014tka,QCDev}. 
It turns out that at the Higgs mass scale $M_H$ of about 125 GeV, the effects of linear gluon polarization are not expected to be large, at the few percent level most likely. Moreover, the effects are largest at small values of the transverse momentum of the Higgs, i.e.\ a few GeV, where 
the cross section is difficult to measure. In the present paper we consider an alternative offered by the production of a Higgs in association with an additional jet, which has been widely studied without including gluon polarization, e.g.\ 
Refs.~\cite{Baur:1989cm,Graudenz:1992pv,Abdullin:1998er,deFlorian:1999zd,Belyaev:2010bc,Jouttenus:2013hs,Liu:2013hba,Sun:2014lna,Dawson:2014ora}.
The effect of gluon polarization shows up in the transverse momentum distribution of the Higgs plus jet pair, where the pair transverse momentum can be of the order of a few GeV, while the separate transverse momenta of the Higgs and the jet can be substantially larger. 
The invariant mass of the Higgs plus jet system will be even larger than $M_H$, but the advantage is that a range of scales is now accessible, as opposed to the very narrow range around $M_H$ accessible in Higgs production. In principle, this range of scales offers a way to map out the TMD evolution, although the feasibility in practice remains to be seen. 

Experimentally the limiting factors are the resolution of the transverse momentum of both the Higgs and the jet and how well the jet direction coincides with that of the fragmenting parton. At CMS the jet transverse momentum resolution at 10 GeV is typically 1.5 GeV and at 100 GeV it is 8 GeV \cite{Khachatryan:2014uwa}. The ultimately achievable resolution on the Higgs transverse momentum is not clear, but it is likely multiple GeV. On an event-by-event basis the deviation of the jet axis as obtained by jet finding algorithms from the direction of the fragmenting parton, can also be as large as a few GeV in transverse momentum \cite{CMS:2009dwa}. Altogether these uncertainties in the pair transverse momentum can be substantial and the goal of obtaining several bins in the region up to say 10 GeV will be quite challenging. A numerical study using a Monte Carlo simulation will have to be done to study the actual feasibility, but that is beyond the scope of this paper. Here we focus on the cross section expressions and on the differences of the Higgs plus jet process to Higgs production and to some other similar processes, pointing out the advantages it in principle has to offer. 

In this paper we present the relevant expressions for Higgs plus jet production in leading order and study the impact of the gluon polarization in two models for the gluon distributions involved. Both models have the advantage that they allow to obtain analytic expressions, but we will mostly present numerical results to show the qualitative differences between the two cases more clearly. We also present results for angular distributions, which have the advantage of singling out specific contributions. Although measurements of angular distributions generally require large statistics, probing a nonzero result may nevertheless be possible when integrating over transverse momenta up to some maximum value as suggested in Ref.~\cite{Dunnen:2014eta}.

\section{Outline of the calculation}

\begin{figure}[t]
 \vspace{-2cm}
\centerline{ \epsfig{file=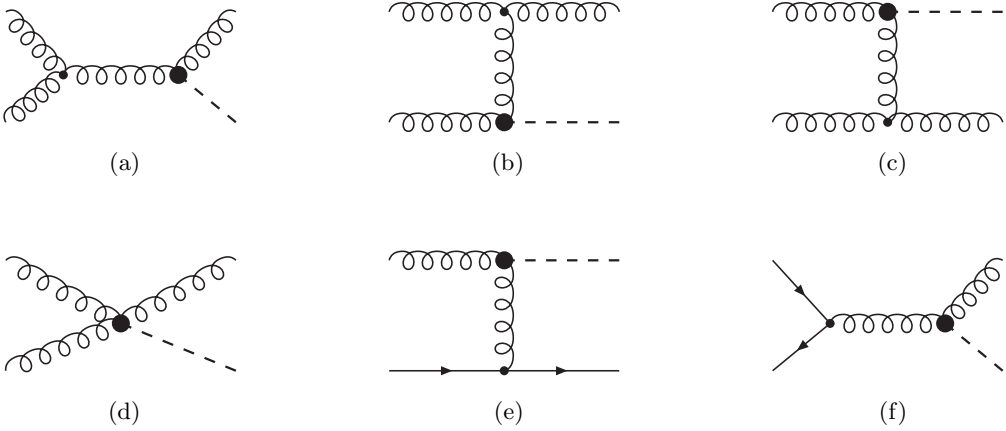}}
  \vspace{-19.5cm}
\caption{\it Feynman diagrams for the partonic subprocesses contributing to 
$p\,p\to H \, {\rm jet}\,X$ at leading order in perturbative QCD: $gg\to Hg$ (a)-(d), $gq\to Hq$ (e), $q \bar{q}\to Hg$ (f). }
\label{fig:fd}
\end{figure}

We study the process 
\begin{equation}
p(P_A)\,{+}\,p(P_B)\,\to\, H (K_H) \,{+}\, {\rm jet} (K_{\rm j})\, {+}  \,X \, ,
\label{eq:proc}
\end{equation}
where the four-momenta of the particles are given within round brackets, and the Higgs boson and jet in the final state are produced with momenta that have 
components in the plane orthogonal to the direction of the initial protons that are almost back to back. To leading order  
in perturbative QCD the reaction proceeds via the partonic subprocesses
\begin{equation}
a(p_a)\,{+}\,b(p_b)\,\to\, H (K_H)\,{+}\, c(K_{\rm j}) \, ,
\label{eq:subproc}
\end{equation}
with parton $c$ fragmenting into the observed jet. Specifically, the following channels can contribute: 
$gg\to Hg$, $gq\to Hq$ and $q \bar q \to Hg $~\cite{Ellis:1987xu,Dawson:1990zj,Kauffman:1996ix}. The corresponding Feynman diagrams are depicted in Fig.~\ref{fig:fd}. In the calculation of the scattering amplitudes, we take the quark masses to be zero, except for the top quark mass $M_t$. Therefore the Higgs boson can couple to gluons only via a top quark loop. We consider the limit $M_t\to\infty$ in which this coupling can be approximated by a point interaction. The corresponding Feynman rules of the effective Lagrangian can be found, for example, in Ref.~\cite{Kauffman:1996ix}. Furthermore, we perform a lightcone decomposition of the two incoming hadronic momenta, $P_A$ and $P_B$, in terms of the light-like vectors $n_+$ and $n_-$, which satisfy the relations $n_+^2\,{=}\,n_-^2\,{=}\,0$ and $n_+{\cdot}n_-\,{=}\,1$:
\begin{equation}
P_A^\mu
=P_A^+n_+^\mu+\frac{M_p^2}{2P_A^+}n_-^\mu\ ,\qquad\text{and}\qquad
P_B^\mu
=\frac{M_p^2}{2P_B^-}n_+^\mu+P_B^-n_-^\mu\ ~.
\end{equation}
The partonic momenta $p_a$ and  $p_b$ can be expressed in terms of the  
lightcone momentum fractions ($x_a$, $x_b$) and the 
intrinsic transverse momenta ($ p_{a \sT}$, $ p_{b \sT}$), as follows
\begin{equation}
p_a^\mu
=x_a^{\phantom{+}}\!P_A^+n_+^\mu
+\frac{p_a^2{+}\boldsymbol p_{a \sT}^2}{2\,x_a^{\phantom{+}}\!P_A^+}n_-^\mu
+p_{a \sT}^\mu\ ,
\qquad\text{and}\qquad
p_b^\mu
=\frac{p_b^2{+}\boldsymbol p_{b \sT}^2}{2\,x_b^{\phantom{-}}\!P_B^-}n_+^\mu
+x_b^{\phantom{-}}\!P_B^-n_-^\mu+p_{b \sT}^\mu\ .
\label{PartonDecompositions}
\end{equation}
Using $n_+$ and $n_-$ the lightcone components of any
vector $v$ are defined as $v^\pm \equiv v \cdot n_\mp$, while $v_\perp$  
refers to the components of  
 $v$ orthogonal to the proton momenta $P_A$ and $P_B$. Moreover, one has 
$v_\perp^2 =-\bm v^2_\perp$. 
Therefore in Eq.~(\ref{PartonDecompositions}), if we neglect the proton mass, 
$p_{a \sT}^{\mu}= p_{a \perp}^{\mu}$
and $p_{b \sT}^{\mu}= p_{b \perp}^{\mu}$.

We assume that, at sufficiently high energies, TMD factorization \cite{Collins:2011zzd,GarciaEchevarria:2011rb} holds for the process in Eq.~(\ref{eq:proc}), hence 
its cross section is given by the convolution of one soft, partonic correlator for each proton and a hard part,
\begin{eqnarray}
\d\sigma
& = &\frac{1}{2 s}\,\frac{d^3 \bm K_H}{(2\pi)^3\,2 E_{H}}\, \frac{d^3 \bm K_{\rm j}}{(2\pi)^3\,2 E_{\rm j}}\,\sum_{a,b,c} 
{\int} \d x_a \,\d x_b \,\d^2\bm p_{a\sT} \,\d^2\bm p_{b\sT}\,(2\pi)^4
\delta^4(p_a{+} p_b {-}K_H- K_{\rm j})
 \nonumber \\
&&\qquad \qquad\qquad \qquad\qquad\times
{\rm Tr}\, \left \{ \Phi^{[U]}_a(x_a {,}\bm p_{a \sT}) \Phi^{[U]}_b(x_b {,}\bm p_{b \sT})
 \left|{\cal M}^{a b  \to H c} (p_a, p_b; K_H, K_{\rm j})\right|^2\right \}\,,
\label{CrossSec}
\end{eqnarray}
with $s = (P_A + P_B)^2$ being the total energy squared in the hadronic 
center-of-mass frame. The sum in Eq.~(\ref{CrossSec}) runs over all the partons
 that take part in the reaction, the appropriate trace is taken over  Dirac and Lorentz indices, and ${\cal M}^{a b  \to H c}$ denotes the amplitude for the 
process $ab\to Hc$. The parton correlators $\Phi^{[U]}_{a,b}$ describe the hadron $\rightarrow$ parton transitions. They can be parameterized in terms of TMDs 
and are defined in terms of QCD operators on the lightfront (LF): $\xi{\cdot}n\,{\equiv}\,0$, where $n\equiv n_-$ for parton $a$ with momentum $p=p_a$ and $n\equiv n_+$ for parton $b$ with momentum $p=p_b$.
Specifically, at leading twist the quark correlator for an unpolarized hadron can be written as~\cite{Boer:1997nt,Bacchetta:2006tn}
\begin{eqnarray}
\label{QuarkCorr}
{\Phi_{q\, ij}^{[U]}}(x{,} \bm p_\sT)
 =  {\int}\frac{\d(\xi{\cdot}P)\,\d^2\xi_\sT}{(2\pi)^3}\ e^{ip\cdot\xi}\,
\langle P |\,\overline\psi_j(0)\,U_{[0,\xi]}\,
\psi_i(\xi)\,|P\rangle\,\big\rfloor_{\text{LF}}  =  \frac{1}{2}\,
\bigg \{\,f_1^q(x{,}\bm{p}_\sT^2)\;\slash P_{ij}
+i h_1^{\perp\,q}(x{,}\bm{p}_\sT^2)\;\frac{[\slash p_\sT , 
\slash P]_{ij}}{2 M_p}\bigg \} , 
\end{eqnarray}
where $U_{[0,\xi]}$ is the process dependent gauge link connecting the 
two quark fields, which renders the correlator gauge invariant. Furthermore, $f_1^q(x, \bm{p}_\sT^2)$ is the TMD describing unpolarized quarks inside an unpolarized hadron, and  $h_1^{\perp q}(x, \bm{p}_\sT^2)$, commonly referred to as the Boer-Mulders function, is the time-reversal ($T$) odd distribution of transversely polarized quarks inside an unpolarized hadron~\cite{Boer:1997nt}. Analogously, for an antiquark one has
\begin{eqnarray}
\label{AquarkCorr}
{{\overline{\Phi}}_{q\, ij}^{[U]}}(x{,} \bm p_\sT)
 =  -{\int}\frac{\d(\xi{\cdot}P)\,\d^2\xi_\sT}{(2\pi)^3}\ e^{-ip\cdot\xi}\,
\langle P |\,\overline\psi_j(0)\,U_{[0,\xi]}\,
\psi_i(\xi)\,|P\rangle\,\big\rfloor_{\text{LF}} 
 =  \frac{1}{2}\,
\bigg \{\,f_1^{\bar q}(x{,}\bm{p}_\sT^2)\;\slash P_{ij}
+i h_1^{\perp\,\bar q}(x{,}\bm{p}_\sT^2)\;\frac{[\slash p_\sT , 
\slash P]_{ij}}{2 M_p}\bigg \}\, . 
\end{eqnarray}
The definition of the gluon correlator in terms of the gluon field strength $F^{\mu\nu}$ has been given for the first time in Ref.~\cite{Mulders:2000sh}. For an unpolarized hadron, using the naming convention of Ref.~\cite{Meissner:2007rx}, 
one has
\begin{eqnarray}
\label{GluonCorr}
\Phi_g^{[U]\,\mu\nu}(x,\bm p_\sT )
& = &  \frac{n_\rho\,n_\sigma}{(p{\cdot}n)^2}
{\int}\frac{\d(\xi{\cdot}P)\,\d^2\xi_\sT}{(2\pi)^3}\
e^{ip\cdot\xi}\,
\langle P|\,\tr\big[\,F^{\mu\rho}(0)\,U_{[0,\xi]}\,
F^{\nu\sigma}(\xi)\,U^{\prime}_{[\xi,0]}\,\big]
\,|P \rangle\,\big\rfloor_{\text{LF}} \nonumber \\
& =& 
-\frac{1}{2x}\,\bigg \{g_\sT^{\mu\nu}\,f_1^g (x,\bm p_\sT^2)
-\bigg(\frac{p_\sT^\mu p_\sT^\nu}{M_p^2}\,
{+}\,g_\sT^{\mu\nu}\frac{\bm p_\sT^2}{2M_p^2}\bigg)
\;h_1^{\perp\,g} (x,\bm p_\sT^2) \bigg \} , \label{Phig}
\end{eqnarray}
with the transverse tensor $g^{\mu\nu}_\sT$ defined as $g^{\mu\nu}_{\sT} = g^{\mu\nu}
- n_+^{\mu}n_-^{\nu} -n_-^{\mu}n_+^{\nu}$. The function
$f_1^g(x,\bm{p}_\sT^2)$ is the unpolarized gluon distribution and 
$h_1^{\perp\,g}(x,\bm{p}_\sT^2)$ the distribution of linearly polarized gluons, which satisfies the model-independent positivity bound~\cite{Mulders:2000sh},
\begin{equation}
\frac{\bm p_\sT^2}{2M_p^2}\,|h_1^{\perp \,g}(x,\bm p_\sT^2)|\le f_1^g(x,\bm p_\sT^2)\,,\label{eq:Bound}
\end{equation}
valid for all values of $x$ and $\bm p_\sT$. In contrast to $h_1^{\perp\, q}(x, \bm p^2_\sT)$,  $h_1^{\perp\, g}(x, \bm p_\sT^2)$ is $T$-even, hence it can be 
nonzero also in absence of initial and/or final state interactions. However, 
as any other TMD, $h_1^{\perp\, g}(x, \bm p_\sT^2)$ 
can in principle receive contributions from these interactions, which can 
render it process-dependent and even hamper its extraction for processes where factorization does not hold, such as dijet production in hadron-hadron collisions~\cite{Boer:2009nc,Rogers:2010dm,Buffing:2011mj}. 

The  cross section is calculated using Eq.~(\ref{CrossSec}), in which 
we insert the parameterizations of the correlators in Eqs.~(\ref{QuarkCorr})-(\ref{GluonCorr}) and the explicit expressions of the amplitudes ${\cal M}^{ab\to Hc}$. From the $\delta$ function in Eq.~(\ref{CrossSec}), with $p_a^-=p_b^+ =0$, 
the lightcone momentum fractions $x_{a,b}$ can be expressed in terms of the rapidities ($y_H$, $y_{\rm j}$) and transverse momenta ($\bm K_{H \perp}$, $\bm K_{{\rm j} \perp}$) of the produced Higgs boson and jet, respectively:
\begin{equation} 
x_a \, =\, \frac{M_{\perp}\,e^{y_H} \, + \, \vert \bm K_{{\rm j} \perp} \vert \,
e^{y_{\rm j}}}{\sqrt s}, \,\qquad
x_b \, {=}\, \frac{M_{\perp}\,e^{-y_H} \, + \, \vert \bm K_{{\rm j} \perp} \vert \,
e^{-y_{\rm j}}}{\sqrt s} \, , 
\end{equation} 
valid with corrections of order ${\cal O}(1/s)$ and where we have introduced 
the transverse mass $M_\perp = \sqrt{M_H^2 + \bm K_{H \perp}^2} \geq M_H$. If we neglect terms suppressed by powers of $\vert \bm q_\sT \vert /  M_\perp $, the final result has the form 
\begin{equation}
\frac{\d\sigma}
{\d y_H \, \d y_{\rm j} \,\d^2 \bm{K}_{\perp} \,\d^2 \bm{q}_{\sT}} =
\frac {\alpha_s^3}{144\, \pi^3 \,v^2}\,\frac{1}{x_ax_b s^2}\,
\bigg[ A(\bm{q}_\sT^2) + B(\bm{q}_\sT^2) \cos 2 \phi + 
C(\bm{q}_\sT^2) \cos 4 \phi \bigg ]\, ,
\label{eq:cs}
\end{equation}
where we have introduced the sum and difference of the final transverse 
momenta, $\bm K_\perp = (\bm K_{H \perp} - \bm K_{{\rm j} \perp})/2$ and 
$\bm q_\sT = \bm K_{H \perp} + \bm K_{{\rm j} \perp}$. Moreover, $v$ is the vacuum 
expectation value and $\phi$ is the azimuthal angle between  $\bm K_\perp$ and $\bm q_\sT$, namely $ \phi =  \phi_\perp- \phi_\sT$. The functions 
$A$, $B$, and $C$ contain convolutions of the various TMDs and, besides $\bm q_{\sT}^2$, 
they depend also on the Mandelstam variables $\hat s$, $\hat t$, $\hat u$ for the partonic subprocesses in Eq.~(\ref{eq:subproc}), which 
satisfy the relations
\begin{eqnarray}
\hat s & = & (p_a + p_b)^2\, = \, 2\, p_a\cdot p_b \,= \,  (K_H+K_{\rm j})^2 \, = M_H^2 + 2 \,K_H \cdot K_{\rm j}\, = \, x_a x_b s \,, \nonumber \\
\hat t  & = &   (p_a-K_H)^2\, = \,   M_H^2 -2 \, p_a \cdot K_H \,= \,  M_H^2 - x_a \,M_\perp \,\sqrt{s}\, e^{-y_H} \nonumber \\
& = & (p_b - K_{\rm j})^2\, = \,-2 \,p_b\cdot K_{\rm j}\, = \,  -x_b \,\vert \bm K_{{\rm j} \perp}\vert  \,\sqrt{s}\, e^{y_{\rm j}}        \, , \nonumber \\
\hat u  & = &  (p_a-K_{\rm j})^2  \,=\,  -2\, p_a \cdot  K_{\rm j} \, = \,-x_a \,\vert \bm K_{{\rm j} \perp}\vert  \,\sqrt{s}\, e^{-y_{\rm j}} \nonumber \\
& = &  (p_b-K_H)^2\, = \,M_H^2-2\, p_b\cdot K_H\,  = \, M_H^2 -x_b \,M_\perp \,\sqrt{s}\, e^{y_H}\,, 
\end{eqnarray}
with  $\hat s + \hat t + \hat u = M_H^2$ . The explicit expressions for $A$, $B$ and $C$ are provided in the following three subsections. 

\subsection{Angular independent part of the cross section}

The term $A$ in Eq.~(\ref{eq:cs}) is given by the 
sum of contributions from the relevant partonic subprocesses, i.e.\ 
\begin{equation}
A (\bm q_\sT^2) = \sum_{a,b,c} {\cal A}_f^{ab\to Hc}\,+ \, {\cal A}_h^{gg\to Hg}\,,
\label{eq:A}
\end{equation}
where
\begin{eqnarray}
{\cal A}_f^{gg \to Hg} & = & \frac{N_c}{N_c^2-1}\,\frac{M_H^8 + \hat s^4 + \hat t^4 + \hat u^4}{\hat s \hat t \hat u} \, {\cal C}[f_1^g\,f_1^g]\,,\label{eq:Afgg}\\
{\cal A}_f^{gq \to Hq} & = & \frac{1}{2 N_c}
\left \{ - \frac{\hat s^2 + \hat u^2}{\hat t} \,{\cal C}[f_1^g\,f_1^q] -  \frac{\hat s^2 + \hat t^2}{\hat u} \,
 {\cal C}[f_1^q\,f_1^g] \right \}\,,\\
{\cal A}_f^{q\bar q \to Hg} & = &  \frac{N_c^2-1 }{2 N_c^2} \,
 \frac{\hat t^2 + \hat u^2}{\hat s} \,  {\cal C}[f_1^q\,f_1^{\bar q}] \label{eq:Afqq}\, ,\\
 {\cal A}_h^{gg\to Hg} & = &  \frac{N_c}{N_c^2-1}\,M_H^4\, \frac{\hat s}{\hat t \hat u} \,{\cal C}[w_0^{hh}\,h_1^{\perp\,g}h_1^{\perp\,g}] =
 \frac{1}{9}\,\frac{N_c}{N_c^2-1}\,\frac{M_H^4}{\bm K_\perp^2} \,{\cal C}[w_0^{hh}\,h_1^{\perp\,g}h_1^{\perp\,g}]\,. \label{eq:Ahgg}
\end{eqnarray}
In Eqs.~(\ref{eq:Afgg})-(\ref{eq:Afqq}), $N_c$ is the number of colors and we 
have introduced the convolutions of TMDs 
\begin{eqnarray}
{\cal{C}}[w\, f\, f] & \equiv & \int d^{2}\bm p_{a\sT}\int d^{2}\bm p_{b\sT}\,
\delta^{2}(\bm p_{a\sT}+\bm p_{b\sT}-\bm q_{\sT}) \, w(\bm p_{a\sT},\bm p_{b\sT})\, f(x_{a},\bm p_{a\sT}^{2})\, f(x_{b},\bm p_{b\sT}^{2})\,,\label{eq:Conv}
\end{eqnarray} 
with the transverse weight $w_0^{hh}$ given by
\begin{equation}
w_0^{hh} = \frac{1}{M_p^4}\, \left[ (\bm p_{a\sT}\cdot \bm p_{b\sT})^2 - \frac{1}{2}\, \bm p_{a\sT}^2 \,\bm p_{b\sT}^2\right ]\,.
\label{eq:w0}
\end{equation}
The expressions in Eqs.~(\ref{eq:Afgg})-(\ref{eq:Afqq}) are in full agreement with the unpolarized partonic cross sections calculated for the first time in Ref.~\cite{Ellis:1987xu}. The term in Eq.~(\ref{eq:Ahgg}), due to the presence of linearly polarized gluons inside an unpolarized proton, is 
a new result, similar to the modifications of the transverse momentum 
distribution of Higgs bosons~\cite{Boer:2011kf,Boer:2013fca} and (pseudo)scalar 
quarkonia~\cite{Boer:2012bt} inclusively produced in hadronic collisions.

\subsection{The $\cos 2\phi$ angular distribution of the Higgs-jet system}

Similarly to Eq.~(\ref{eq:A}), the term $B$ in Eq.~(\ref{eq:cs}) can be 
written as
\begin{equation}
B(\bm q_{\sT}^2)  = \sum_{a,b,c} {\cal B}^{ab\to Hc}\,,
\end{equation}
where
\begin{eqnarray}
 {\cal B}^{gg\to Hg} & = & \frac{N_c}{N_c^2-1}\, \left \{  \frac{ \hat t ^2(\hat t + \hat u)^2 -2 M_H^2 \hat u^2 (\hat t + \hat u) + M_H^4 (\hat t^2 + \hat u^2)  }{\hat s \hat t \hat u}  \, {\cal C}[w_2^{fh} \,f_1^g\,h_1^{\perp\,g}] \right \} + (x_a \leftrightarrow x_b , \hat t \leftrightarrow \hat u )\,,
\label{eq:Bgg}\\
 {\cal B}^{gq\to Hq} & = & \frac{1}{2 N_c}\, \left \{ \frac{\hat s \hat u}{\hat t}\, {\cal C}[w_2^{fh} \,f_1^q\,h_1^{\perp\,g}] +\frac{\hat s \hat t}{\hat u}\, {\cal C}[w_2^{hf} \, h_1^{\perp\,g}\,f_1^q] \right \}\,, \label{eq:Bgq}\\
 {\cal B}^{q\bar q\to H g} & = & \frac{N_c^2-1}{2 N_c}\, \frac{\hat t \hat u}{\hat s}\, 2\,{\cal C}[w_2^{hh} \,h_1^{\perp\, q}\,h_1^{\perp \,\bar q}] \,,
\end{eqnarray}
and the transverse weights read 
\begin{eqnarray}
w_2^{fh} & = & \frac{1}{M_p^2}\, \left [ 2\, 
\frac{(\bm q_\sT \cdot \bm p_{b\sT})^2}{\bm q_\sT^2} -\bm p_{b\sT}^2   \right ] \,,\\
 w_2^{hf}\, & = & \, \frac{1}{M_p^2}\,\left [ 2\, 
\frac{(\bm q_\sT \cdot \bm p_{a\sT})^2}{\bm q_\sT^2}-\bm p_{a\sT}^2   \right ]\,,\\
w_2^{hh} & = & \frac{1}{M_p^2}\,\left [2\,\frac{(\bm q_{\sT} \cdot \bm p_{a\sT})\cdot (\bm q_{\sT} \cdot \bm p_{b\sT})}{\bm q_{\sT}^2} -\bm p_{a\sT}\cdot \bm p_{b\sT}  \right]\,.
\end{eqnarray}
A $\cos 2 \phi$ double Boer-Mulders (quark) contribution with transverse weight $w_2^{hh}$ has been found for the Drell-Yan process as well, and it is expected to lead to a violation of the Lam-Tung relation \cite{Boer:1999mm,Boer:2002ju}. A similar asymmetry has been predicted for photon+jet~\cite{Boer:2007nd} and dijet production~\cite{Lu:2008qu,Boer:2009nc} in proton-proton collision. A $\cos 2 \phi$ modulation due to the convolution of unpolarized and polarized gluon distributions as in Eq.~(\ref{eq:Bgg}) has been predicted also for the inclusive hadroproduction of diphotons~\cite{Qiu:2011ai,Boer:2013fca}, dijets~\cite{Boer:2009nc}, heavy quark pairs~\cite{Boer:2010zf,Pisano:2013cya}, and $J/\psi$+photon pairs~\cite{Dunnen:2014eta}.    

\subsection{The $\cos 4\phi$ angular distribution of the Higgs-jet system}
The only channel that contributes to the $\cos4\phi$ modulation of the cross
section is $gg\to Hg$. Therefore we can write
\begin{eqnarray}
C(\bm q_{\sT}^2) = {\cal C}^{gg\to Hg}  = \frac{N_c}{N_c^2-1}\,\frac{\hat t \hat u}{\hat s} \,{\cal C}[w_4^{hh}\,h_1^{\perp\,g}h_1^{\perp\,g}] = \frac{1}{9}\,\frac{N_c}{N_c^2-1}\,\bm K_\perp^2 \,{\cal C}[w_4^{hh}\,h_1^{\perp\,g}h_1^{\perp\,g}]\,,
\end{eqnarray}
with
\begin{eqnarray}
w_4^{hh} & = & \frac{1}{2 M_p^4}\, \left \{ 2 \,\left [
 2 \, \frac{(\bm q_\sT\cdot \bm p_{a\sT}) (\bm q_\sT\cdot \bm p_{b\sT})   }{\bm q_\sT^2} 
  -\bm p_{a\sT} \cdot  \bm p_{b\sT} \right ]^2 - \bm p_{a\sT}^2  \bm p_{b\sT}^2 \right \}~. 
\end{eqnarray}
An analogous $\cos 4 \phi$ modulation has been predicted for the first time for the inclusive hadroproduction dijets~\cite{Boer:2009nc}, and subsequently also for reactions with diphotons~\cite{Qiu:2011ai,Boer:2013fca}, heavy quark pairs~\cite{Boer:2010zf,Pisano:2013cya}, and $J/\psi$+photon pairs~\cite{Dunnen:2014eta} in the final state.    

\section{Transverse momentum dependent observables}

The (normalized) cross section for the process $p\,p\to H\,{\rm jet}\,X$, differential in $\bm q_\sT^2$ and $\phi$, is defined as 
\begin{equation}
\frac{\d\sigma}{ \sigma} \equiv
\frac{ \d\sigma}
{\int_0^{q_{\sT \rm max }^2} \d \bm q_\sT^2\int_0^{2 \pi} \d\phi\, \d\sigma}\,, 
\label{eq:qTdist}
\end{equation}
where, restricting now only to the subprocess $gg\to Hg$, $\d\sigma$ is given by
\begin{equation}
\d\sigma \equiv  \frac{\d\sigma}{\d y_H\, \d y_{\rm j} \,\d^2 \bm{K}_{\perp} \,\d^2 \bm q_\sT} = \frac {\alpha_s^3}{144\, \pi^3 \,v^2}\,\frac{1}{x_ax_b s^2}\,
\bigg[ A_f^{gg\to Hg} + A_h^{gg\to Hg} + B^{gg\to Hg}\cos 2 \phi + 
C^{gg\to Hg} \cos 4 \phi \bigg ]\,.
\label{eq:csgg}
\end{equation}
By substituting into Eq.~(\ref{eq:qTdist}) the expression for $\d\sigma$ in Eq.~(\ref{eq:csgg}),
one obtains 
\begin{eqnarray}
\frac{\d \sigma}{\sigma} & = & \frac{1}{2\pi}\,\sigma_0(\bm q_\sT^2) \,\left [ 1 + R_0(\bm q_\sT^2) + R_2(\bm q_\sT^2) \cos2\phi +  R_4(\bm q_\sT^2) \cos4\phi\right ]\,,
\label{eq:csgg-2}
\end{eqnarray}
with
\begin{equation}
\sigma_0(\bm q_\sT^2) \equiv \frac{{\cal C}[f_1^g \, f_1^g ]}{\int_0^{{q^2_{\sT \rm max }}} \d \bm q^2_\sT \, {\cal C}[f_1^g \, f_1^g ]}\,,
\label{eq:sigma0}
\end{equation}
and 
\begin{eqnarray}
R_0(\bm q_\sT^2)& =& \frac{M_H^4\,\hat s^2}{M_H^8 + \hat s^4 + \hat t^4 + \hat u^4}\, \frac{{\cal C}[w_0^{hh}\,h_1^{\perp\,g} \, h_1^{\perp\,g} ] } {{\cal C}[f_1^g \, f_1^g ]}\,, \label{eq:R0} \\
R_2(\bm q_\sT^2)& = & \frac{ \hat t ^2(\hat t + \hat u)^2 -2 M_H^2 \hat u^2 (\hat t + \hat u) + M_H^4 (\hat t^2 + \hat u^2)  }{M_H^8 + \hat s^4 + \hat t^4 + 
\hat u^4 }  \, \frac{{\cal C}[w_2^{fh} \,f_1^g\,h_1^{\perp\,g}]}{{\cal C}[f_1^g \, f_1^g ]}  + (x_a \leftrightarrow x_b , \hat t \leftrightarrow \hat u )\,,
\label{eq:R2}\\
R_4(\bm q_\sT^2) & = & \frac{\hat t^2 \hat u^2}{M_H^8+\hat s^4 + \hat t^4 + \hat u^4} \,\frac{{\cal C}[w_4^{hh}\,h_1^{\perp\,g}h_1^{\perp\,g}]}{{\cal C}[f_1^g \, f_1^g ]}\,.
\label{eq:R4}
\end{eqnarray}

It is possible to single out the different terms $1+R_0$, $R_2$, $R_4$ in 
Eq.~(\ref{eq:csgg-2}) by defining the observables~\cite{Dunnen:2014eta}
\begin{equation}
\langle \cos n\phi \rangle_{q_\sT} \equiv 
\frac{\int_0^{2 \pi} \d \phi \,\cos n\phi\, 
\d\sigma}{\sigma}\,,\qquad n=0, 2, 4\,,
\label{eq:cosnphiqT}
\end{equation}
such that the average values of $\cos n\phi$ are given by the integrals of 
 $\langle\cos n\phi\rangle_{q_{\sT}}$ over $\bm q_{\sT}^2$,
\begin{equation}
\langle \cos n\phi \rangle \equiv \frac{\int_0^{q_{\sT \rm max }^2} \d\bm q_{\sT}^2\int_0^{2 \pi} \d \phi \,\cos n\phi\, 
\d\sigma}{\sigma}  = \int_0^{q_{\sT \rm max }^2} \d\bm q_{\sT}^2\,\langle \cos n\phi \rangle_{q_\sT} \,,\qquad n=0, 2, 4\,.
\label{eq:cosnphi}
\end{equation}
We will comment on the value of $q_{\sT \rm max }^2$ in the next section. 

It can be easily shown that
\begin{eqnarray}
\frac{1}{\sigma}\,\frac{\d\sigma}{\d \bm q_{\sT}^2} & \equiv & \langle 1 \rangle_{q_\sT} \,  = \,  \sigma_0(\bm q_\sT^2)\, [ 1+ R_0(\bm q_\sT^2)]\,,
\label{eq:qT} \\
\langle \cos 2\phi \rangle_{q_\sT} & = &  \frac{1}{2}\,
\sigma_0(\bm q_{\sT}^2)\, R_2(\bm q_{\sT}^2)  \,, \\
\langle \cos 4\phi \rangle_{q_\sT} & = &   \frac{1}{2}\,
\sigma_0(\bm q_{\sT}^2)\, R_4(\bm q_{\sT}^2) \,.
\end{eqnarray}
In the next section we provide numerical estimates for these observables in the specific configuration in which the Higgs boson and the jet have the same rapidities ($y_H=y_{\rm j}$). In this particular kinematic domain, the expressions in Eqs.~(\ref{eq:R0})-(\ref{eq:R4}) reduce to   
\begin{eqnarray}
R_0(\bm q_\sT^2) & = & \, \frac{1}{2}\,\frac{M_H^4}{9 \,\bm K^4_\perp + 8\, \bm K_\perp^2\, M_H^2 + M_H^4}\, \frac{{\cal C}[w_0^{hh}\,h_1^{\perp\,g} \, h_1^{\perp\,g} ] } {{\cal C}[f_1^g \, f_1^g ]}\,, \label{eq:R0-2}\\
R_2(\bm q_\sT^2)  & = & \frac{\bm K_\perp^2 (2 \,\bm K_\perp^2 +M_H^2)}{9 \,\bm K^4_\perp + 8\, \bm K_\perp^2\, M_H^2 + M_H^4}\,  
 \frac{  {\cal C}[w_2^{fh} \,f_1^{g}\, h_1^{\perp\,g}] +  {\cal C}[w_2^{hf} \,h_1^{\perp\,g}\,f_1^g] }{{\cal C}[f_1^g\,f_1^g]}\, ,\\
 R_4(\bm q_\sT^2) &  = & \frac{1}{2}\,\frac{\bm K_\perp^4}{9 \,\bm K^4_\perp + 8\, \bm K_\perp^2\, M_H^2 + M_H^4} \,\frac{ {\cal C}[w_4^{hh} \,h_1^{\perp\,g}\, h_1^{\perp\,g}]}{{\cal C}[f_1^g\,f_1^g]}\,.
 \label{eq:R4-2}
\end{eqnarray} 

\section{Numerical study}

\subsection{Gaussian+tail model}
In order to quantify the effects of gluon polarization on the observables defined in the previous section, we assume that the unpolarized gluon TMD is {\it approximately} a Gaussian at small transverse momentum, but has the proper power law fall-off at large transverse momentum~\cite{Boer:2014tka}:
\begin{equation}
f_1^g(x,\bm p_\sT^2) = {f_1^g(x)}\, \frac{R^2}{2\,\pi}\, \frac{1}{1 + \bm p_\sT^2\,R^2}\,,
\label{eq:f1gdb}
\end{equation}
where $f_1^g(x)$ is the gluon distribution integrated over the transverse 
momentum squared $\bm p_{\sT}^2$ and we choose $R= 2$ GeV$^{-1}$. In order to show the maximal effects, for the distribution of linearly polarized gluons we take 
\begin{equation}
h_1^{\perp \,g}(x,\bm p_\sT^2)=\frac{2M_p^2}{\pT^2} f_1^g(x,\pT^2)\,,
\label{eq:h1pg}
\end{equation}
with $f_1^g$  given by Eq.~(\ref{eq:f1gdb}). In this case the bound in Eq.~(\ref{eq:Bound}) is saturated for every value of $x$ and $\bm p_\sT^2$. In analogy to Eq.~(\ref{eq:f1gdb}), we also consider the 
following model from Ref.\ \cite{Boer:2014tka},  
\begin{equation}
h_1^{\perp\,g}(x,\bm p_\sT^2) = c\,f_1^g(x)\, \frac{M^2_pR_h^4}{2\,\pi}\,
\frac{1}{(1+\bm p_\sT^2R_h^2)^2}\,,
\label{eq:h1pgdb}
\end{equation}
with $c= \pm 2$ and $R_h = 3 R/2$, for which the bound is saturated only in 
the limit $p_\sT \to \infty$.

\begin{figure}[t]
\begin{center}
\includegraphics[width=7.cm]{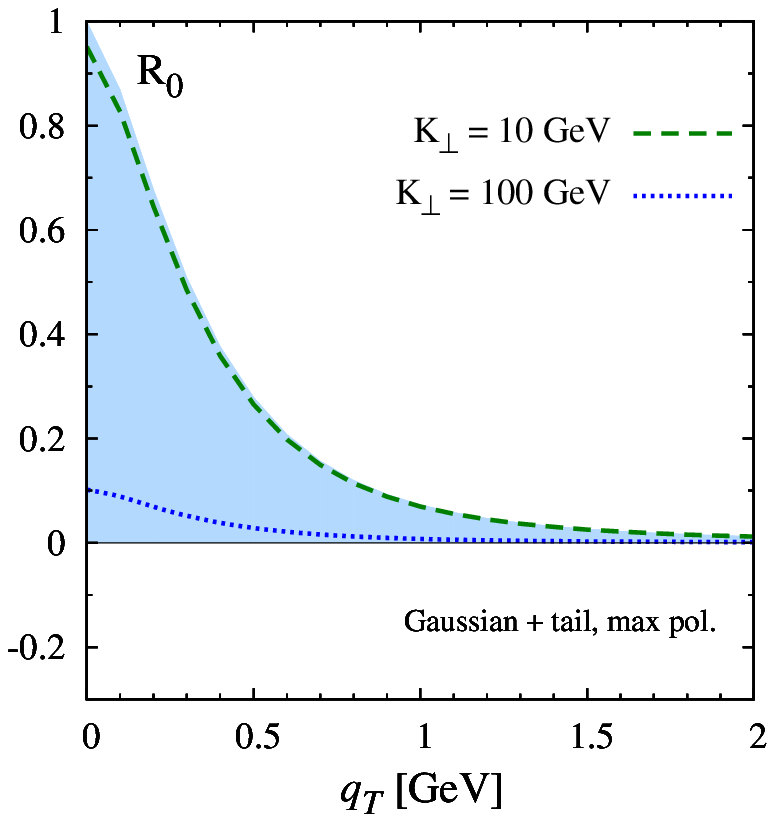}
\includegraphics[width=7.cm]{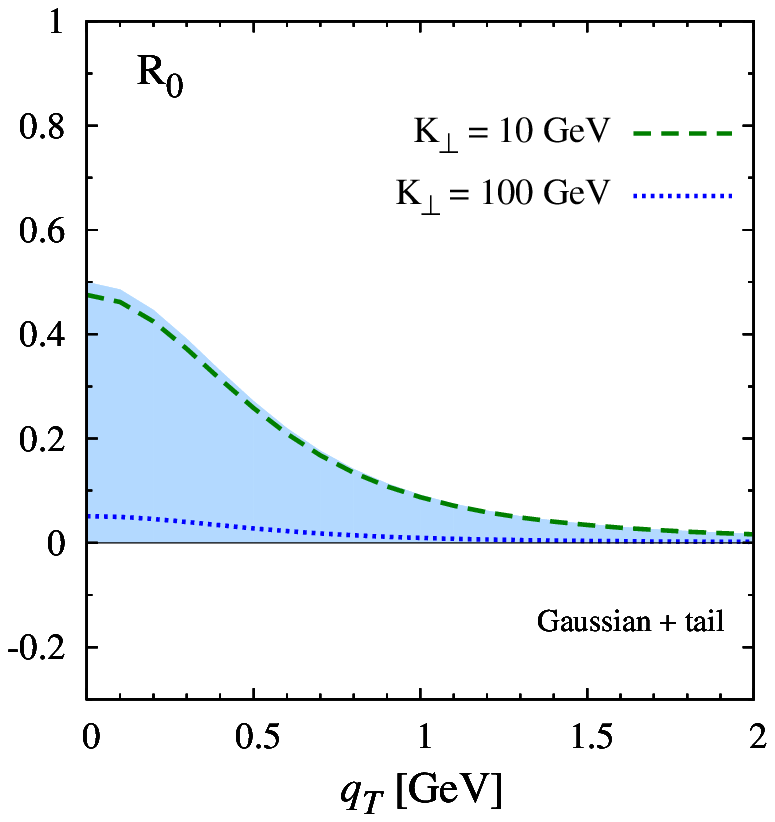}
\end{center}
\caption{\it Upper bounds of $R_0$ defined in Eq.~(\ref{eq:R0}) for the subprocess $gg\to Hg$ and  $y_H=y_{\rm j}$, see also Eq.~(\ref{eq:R0-2}). The unpolarized distribution $f_1^g$ is taken as in Eq.~(\ref{eq:f1gdb}), while $h_1^{\perp\,g}$ either saturates its positivity bound in Eq.~(\ref{eq:h1pg}) (left panel), or it is given by Eq.~(\ref{eq:h1pgdb}) (right panel). The blue band indicates the range for $K_\perp \to 0$. }
\label{fig:R0_db}
\end{figure}

It is convenient to consider the Fourier transforms of the above functions in Eqs.~(\ref{eq:f1gdb}) and (\ref{eq:h1pgdb}):
\begin{eqnarray}
\tilde{f}_1^{g}(x,b^2) &  =  & \int d^2\bm p_T^{}\; e^{-i \bm{b} \cdot \bm{p}_T^{}}\; f_1^{g}(x,p_T^2)= 
f_1^g(x) K_0(b/R),\\
\tilde{h}_1^{\perp\,g}(x,b^2) & = & \int d^2\bm p_T^{}\; \frac{(\bm{b}\!\cdot \!
\bm p_T^{})^2 - \frac{1}{2}\bm{b}^{2} \bm p_T^{2}}{b^2 M^2}
\; e^{-i \bm{b} \cdot \bm{p}_T^{}}\; h_1^{\perp g}(x,p_T^2) \nonumber \\
& = & -\pi \int dp_T^2 \frac{p_T^2}{2M^2} J_2(bp_T) h_1^{\perp g}(x,p_T^2) = \frac{c}{4} f_1^g(x) \frac{b}{R_h} K_1(b/R_h). 
\end{eqnarray}
In this way the relevant convolutions can be expressed as:
\begin{eqnarray} 
& & \mathcal{C}\left[f_{1}^{g}\, f_{1}^{g}\right] =   f_1^g(x_a) f_1^g(x_b)  \int_0^\infty \frac{db}{2\pi}b J_0(b|\bm q_\sT|)  K_0(b/R)^2 ,\\
& & \mathcal{C}\left[w_0^{hh}\,h_1^{\perp\,g}h_1^{\perp\,g} \right] =  \frac{c^2}{8} f_1^g(x_a) f_1^g(x_b) \int_0^\infty \frac{db}{2\pi}b J_0(b|\bm q_\sT|) \frac{b^2}{R_h^2} K_1(b/R_h)^2 ,\\
& & \mathcal{C}\left[w_2^{fh}\,f_1^{g}h_1^{\perp\,g} \right]  =  - \frac{c}{2} f_1^g(x_a) f_1^g(x_b)  \int_0^\infty\frac{db}{2\pi}b J_2(b|\bm q_\sT|) \frac{b}{R_h} K_0(b/R) K_1(b/R_h) ,\\
& & \mathcal{C}\left[w_4^{hh}\,h_1^{\perp\,g}h_1^{\perp\,g} \right]  =  \frac{c^2}{8} f_1^g(x_a) f_1^g(x_b)  \int_0^\infty \frac{db}{2\pi}b J_4(b|\bm q_\sT|) \frac{b^2}{R_h^2} K_1(b/R_h)^2 .
\end{eqnarray}
By substitution into Eqs.~(\ref{eq:R0-2})-(\ref{eq:R4-2}), one finds 
\begin{eqnarray} 
R_0(\bm q_\sT^2) & = & \frac{1}{2}\,\frac{M_H^4}{9 \,\bm K^4_\perp + 8\, \bm K_\perp^2\, M_H^2 + M_H^4}\, 
\frac{c^2}{8}\frac{\int_0^\infty db b J_0(b|\bm q_\sT|) \frac{b^2}{R_h^2} K_1(b/R_h)^2}{\int_0^\infty db b J_0(b|\bm q_\sT|)  K_0(b/R)^2} ,\\
R_2(\bm q_\sT^2) & = & 
\frac{\bm K_\perp^2 (2 \,\bm K_\perp^2 +M_H^2)}{9 \,\bm K^4_\perp + 8\, \bm K_\perp^2\, M_H^2 + M_H^4}\,  \frac{-c}{2}  \frac{\int_0^\infty db b J_2(b|\bm q_\sT|) \frac{b}{R_h} K_0(b/R) K_1(b/R_h)}{\int_0^\infty db b J_0(b|\bm q_\sT|)  K_0(b/R)^2} ,\\
R_4(\bm q_\sT^2) & = & \frac{1}{4}\,\frac{\bm K_\perp^4}{9 \,\bm K^4_\perp + 8\, \bm K_\perp^2\, M_H^2 + M_H^4} \, \frac{c^2}{8}\frac{ \int_0^\infty db b J_4(b|\bm q_\sT|) \frac{b^2}{R_h^2} K_1(b/R_h)^2}{\int_0^\infty db b J_0(b|\bm q_\sT|)  K_0(b/R)^2}.
\end{eqnarray}

Results for the upper bound of $R_0$ in Eq.~(\ref{eq:R0-2}) are shown 
in Fig.~\ref{fig:R0_db}, where we use the unpolarized TMD distribution in Eq.~(\ref{eq:f1gdb}), while $h_1^{\perp\,g}$ is given by Eq.~(\ref{eq:h1pg})
in the left panel and by Eq.~(\ref{eq:h1pgdb}) in the right panel. 
The results are presented for two different choices of $K_\perp \equiv \vert \bm K_\perp\vert $: $K_\perp \equiv \vert \bm K_\perp\vert = 10 $ and $100$ GeV. 

\begin{figure}[t]
\begin{center}
\includegraphics[width=7cm]{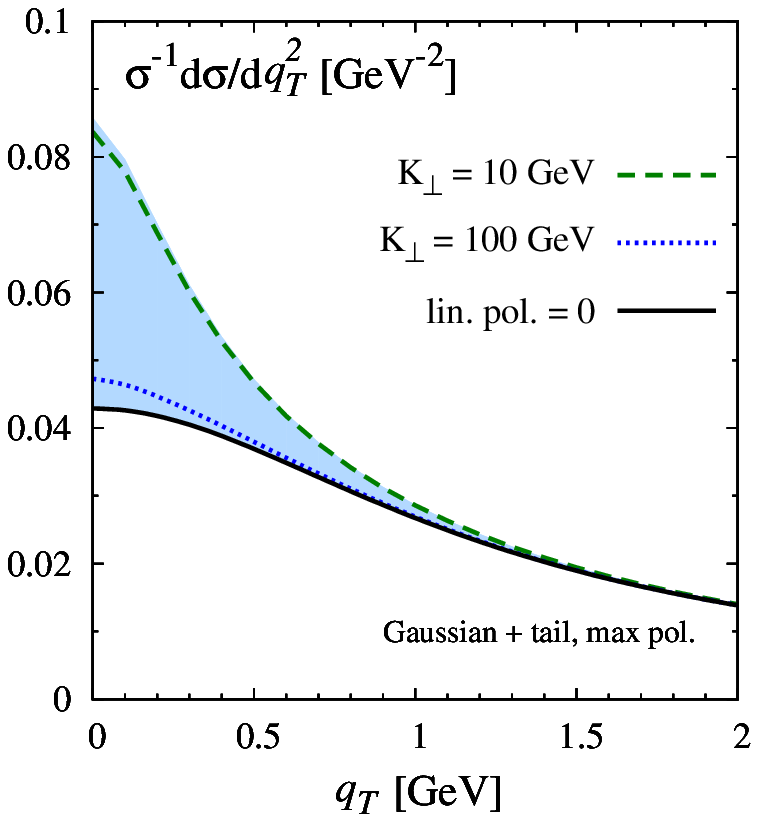}
\includegraphics[width=7cm]{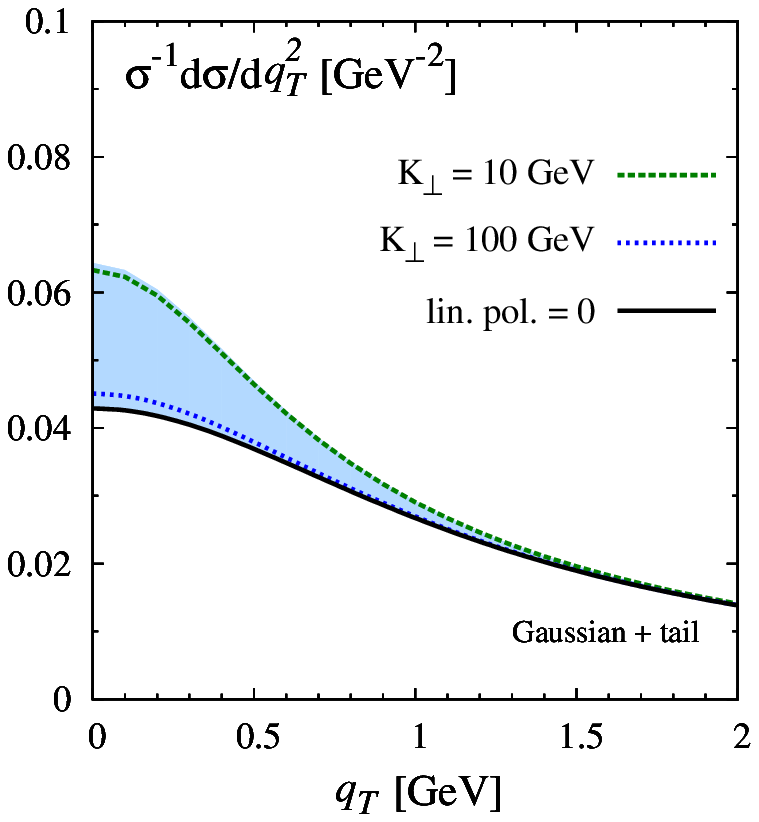}
\end{center}
\caption{\it Transverse momentum distribution of the Higgs plus jet pair in the process $p\,p\to H\,{\rm jet}\, X$, as defined in Eq.~(\ref{eq:qTdist}), for the subprocess $gg\to Hg$, $q_{\sT {\rm max}}^2 =M^2_H/4$, and $y_H=y_{\rm j}$. The TMDs are the same as in Fig.~\ref{fig:R0_db}. 
The solid line indicates the distribution in absence of linear polarization. The blue band indicates the range for $K_\perp \to 0$.}
\label{fig:qT_db}
\end{figure}

The corresponding results for the transverse momentum distribution defined in Eq.~(\ref{eq:qT}), with $\sigma_0$ given in Eq.~(\ref{eq:sigma0}) and 
$q_{\sT {\rm max}}^2 =M^2_H/4$, are depicted in Fig.~\ref{fig:qT_db}. The choice of $q_{\sT {\rm max}}$ is motivated by the requirement of TMD factorization 
that $q_\sT \ll Q$, where $Q$ denotes the hard scale. In the present case we have two hard scales: $M_H$ and $K_\perp$. 
The kinematics considered here is strictly speaking the back-to-back correlation region where $|\bm{q}_\sT| \ll |\bm{K}_\perp |$. 
By integrating up to $q_{\sT {\rm max}}^2 =M^2_H/4$, one however also includes configurations $|\bm{q}_\sT| \simorder |\bm{K}_\perp |$ 
in which $H$ and the jet are not approximately back to back in the lab frame.  
This situation is not included in the calculation of $2 \to 2$ scattering processes presented here. 
However, for the model where the TMD has a power-law tail, the recoil against a third particle emitted into the final state in $2 \to 3$ processes, 
is mimicked to some extent. Differently put, the tail of the TMD is sufficiently hard to produce large-${q}_\sT$ pairs. This is the reason why we 
extend the integration to $q_{\sT {\rm max}}^2 =M^2_H/4$. For the numerical results it does not make too much of a difference. 
In the Gaussian model considered in the next subsection, the tail of the TMDs is too suppressed to mimick the contribution from $2 \to 3$ processes, 
hence, in that case we will restrict to $q_{\sT {\rm max}}^2= K_\perp^2/4$ to emphasize the proper region of validity. 
As a last comment on this point, sometimes the angular distribution of pair production processes are considered in the rest frame of the pair, 
for instance the Collins-Soper frame \cite{Qiu:2011ai,Dunnen:2014eta}. In that case the relative magnitude of $|\bm{q}_\sT|$ w.r.t.\ $|\bm{K}_\perp |$ is not automatically apparent. In the case of Higgs plus jet, the center of mass energy of the pair is generally much larger than $|\bm{q}_\sT|$, 
while $|\bm{q}_\sT|$ can be smaller or larger than $|\bm{K}_\perp |$. If one restricts to $2 \to 2$ scattering processes, one should realize that the region 
$|\bm{q}_\sT| \simorder |\bm{K}_\perp |$ is not properly described, but at best mimicked by including the perturbative tails of the TMDs.

\begin{figure}[t]
\centering
\includegraphics[width=7cm]{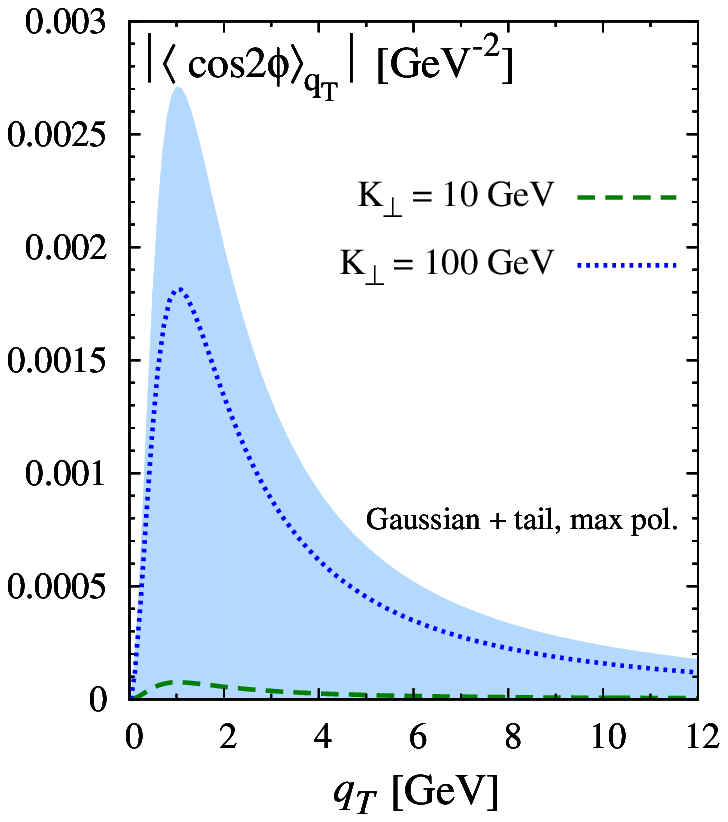}
\includegraphics[width=7cm]{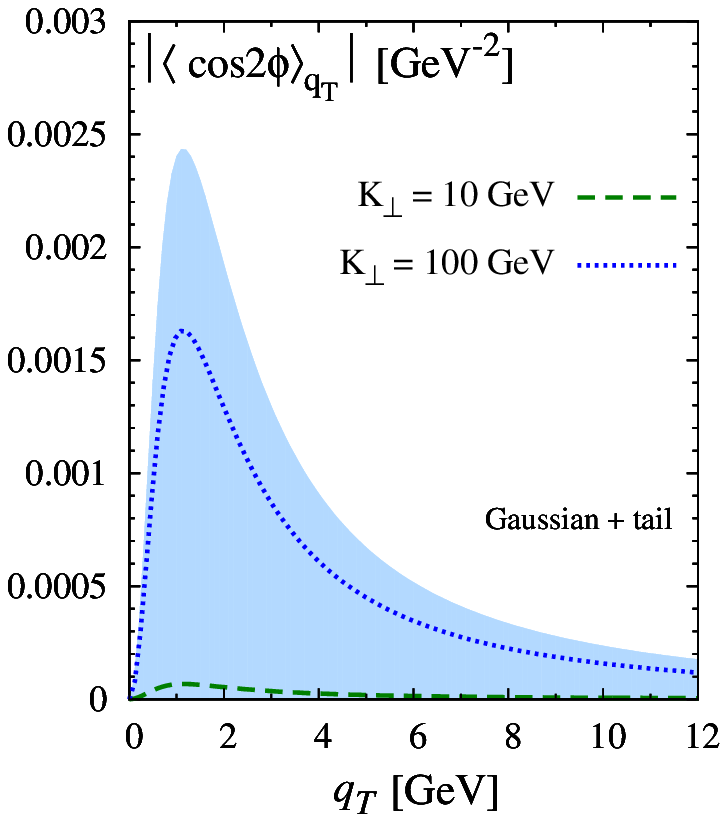}
\caption{\it Absolute value of the $ \langle \cos 2 \phi\rangle_{q_\sT}$  asymmetries for the process  $p\,p\to H \, {\rm jet}\,X$, defined in Eq.~(\ref{eq:cosnphiqT}), as a function of the transverse momentum $q_\sT$ of the Higgs plus jet pair, under the same conditions as in Fig.~\ref{fig:qT_db}. The blue band indicates the range for $K_\perp \to \infty$.}
\label{fig:cos2phi_db}
\end{figure}

\begin{figure}[b]
\centering
\includegraphics[width=7cm]{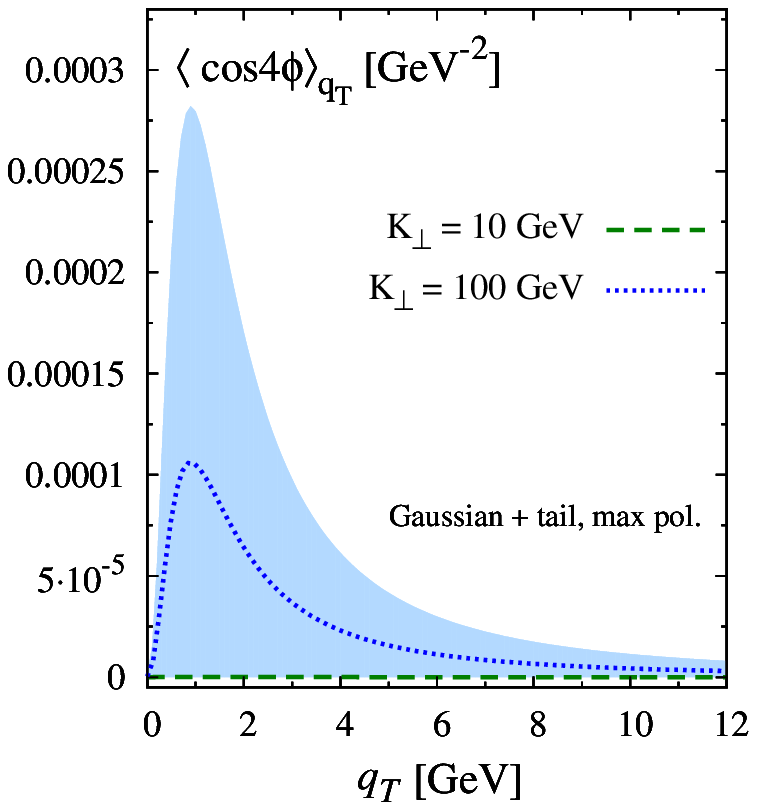}
\includegraphics[width=7cm]{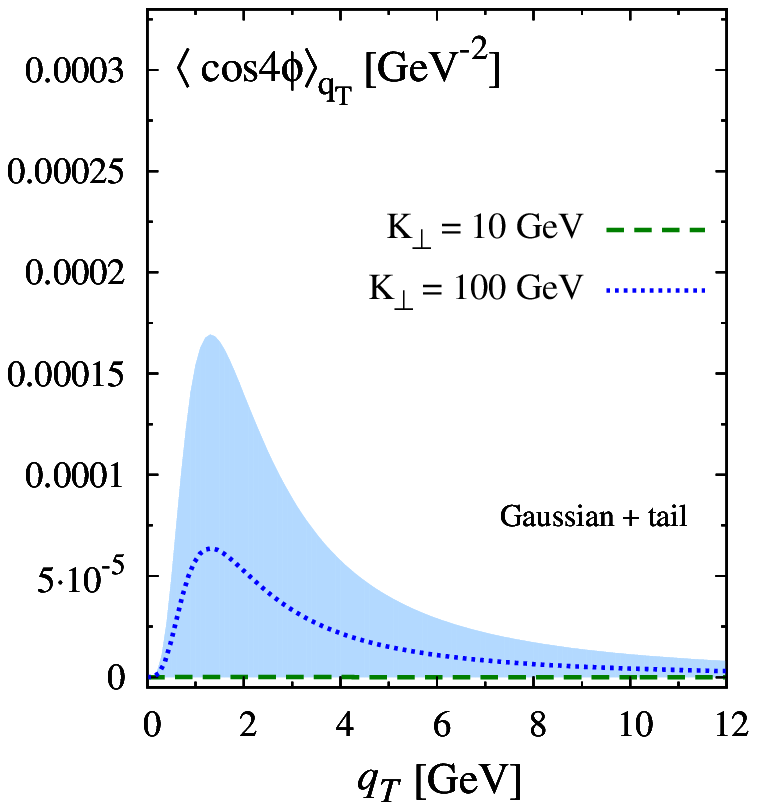}
\caption{\it $\langle \cos 4 \phi\rangle_{q_\sT}$ asymmetries for the process  $p\,p\to H  \, {\rm jet}\,X$, defined in Eq.~(\ref{eq:cosnphiqT}), as a function of the transverse momentum $q_\sT$ of the Higgs plus jet pair, under the same conditions as in Fig.~\ref{fig:qT_db}. The blue band indicates the range for $K_\perp \to \infty$.}
\label{fig:cos4phi_db}
\end{figure}

Our estimates for $\langle \cos 2 \phi\rangle_{q_\sT}$  and $\langle \cos 4 \phi\rangle_{q_\sT}$ are presented in Figs.~\ref{fig:cos2phi_db} and \ref{fig:cos4phi_db}, respectively, with $K_\perp = 10$ and $100$ GeV. As before, for $f_1^g$ we have adopted the {\it Ansatz} in  Eq.~(\ref{eq:f1gdb}),  while  $h_1^{\perp\,g}$  is given either by Eq.~(\ref{eq:h1pg}), in the left panels, or  by Eq.~(\ref{eq:h1pgdb}), in the right panels. Moreover, we have chosen again $q_{\sT {\rm max}}^2 =M^2_H/4$. 


Although we have plotted its absolute value, we point out that $\langle \cos 2 \phi\rangle_{q_\sT}$ 
is the only observable, among the ones discussed here, that is sensitive to 
the sign of the polarized gluon distribution, and it is expected to be
negative if $h_1^{\perp\,g} > 0$. 

Since the magnitudes of $\langle \cos 2 \phi\rangle_{q_\sT}$  and $\langle \cos 4 \phi\rangle_{q_\sT}$  turn out be very small, 
it will be easier to measure the integral of these observables over $\bm q_\sT^2$, up to $q_{\sT {\rm max}}^2$, as defined in Eq.~(\ref{eq:cosnphi}). In both models for $h_1^{\perp\,g}$,  we find that $\vert\langle \cos 2 \phi\rangle\vert \approx 12\%$ when $K_\perp = 100$ GeV, while its value is about 0.5\% when $K_\perp = 10$ GeV. We find that $\langle \cos 4 \phi\rangle$ is about 0.2\% at $K_\perp = 100$ GeV and completely negligible at $K_\perp$ = 10 GeV. These numbers are for 
$q_{\sT {\rm max}}^2 =M^2_H/4$ in both numerator and denominator.

\subsection{Gaussian model}

\begin{figure}[htb]
\centering
{\includegraphics[width=5.5cm]{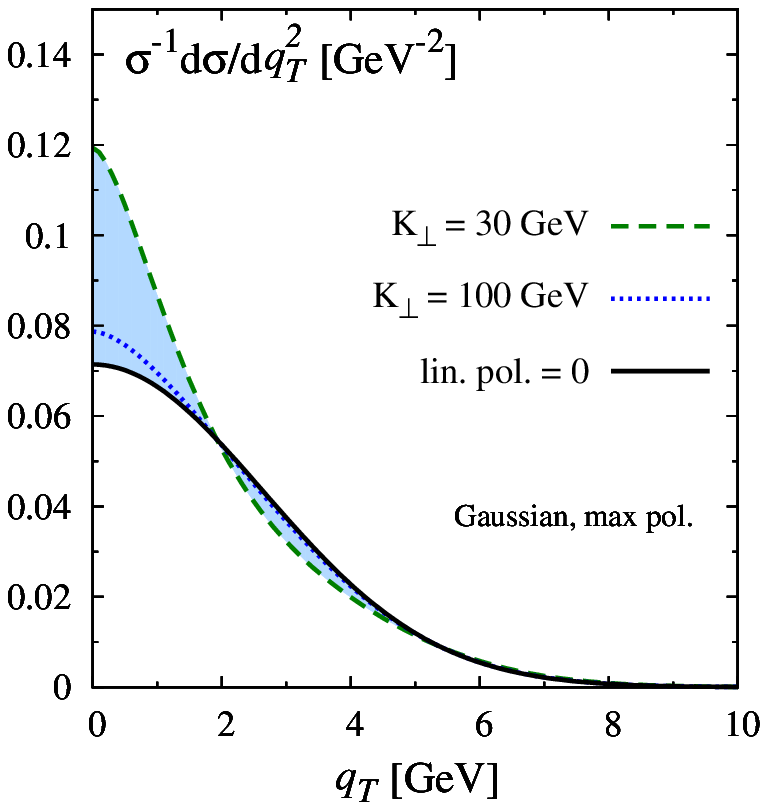}}
{\includegraphics[width=5.5cm]{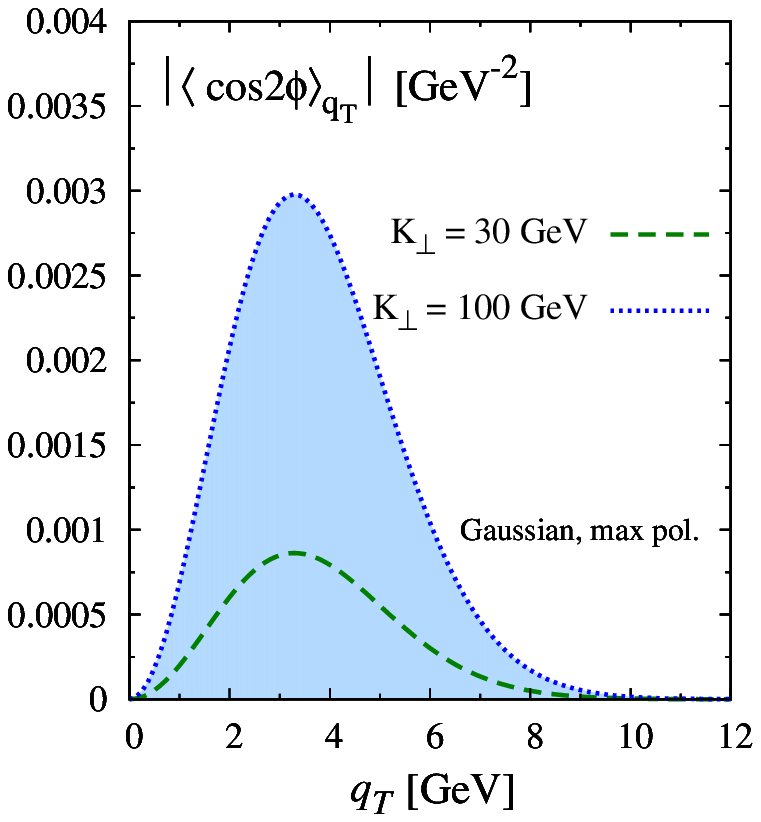}}
{\includegraphics[width=5.5cm]{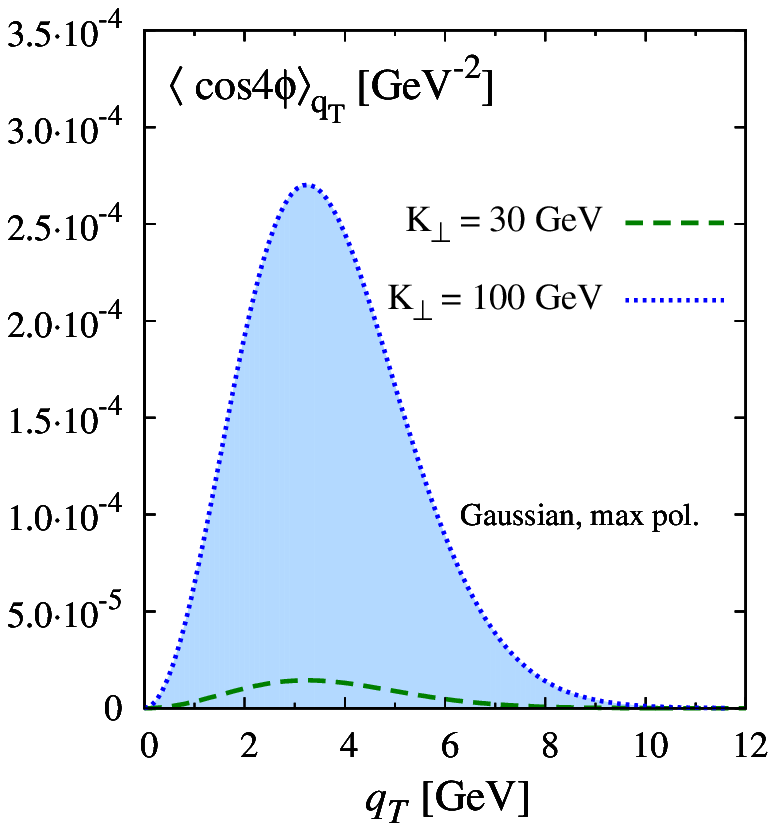}}
\caption{\it Transverse momentum distribution (left panel), $\vert\langle \cos 2 \phi\rangle_{q_{\sT}}\vert$ (central panel) and  $\langle \cos 4 \phi\rangle_{q_\sT}$ (right panel) asymmetries for the process  $p\,p\to H  \, {\rm jet}\,X$, as a function of the transverse momentum $q_\sT$ of the Higgs plus jet pair, in case of equal rapidities and for $q_{\sT {\rm max}}^2= K_\perp^2/4$. The unpolarized distribution $f_1^g$ is taken to have a Gaussian dependence on transverse momentum as in Eq.~(\ref{eq:f1g}) with $\langle p_\sT^2\rangle = 7\ {\rm GeV}^2$, while $h_1^{\perp\,g}$ saturates its positivity bound in Eq.~(\ref{eq:h1pg}).}
\label{fig:gauss}
\end{figure}

For comparison, we now consider a Gaussian model for the TMD distributions, which is widely adopted in many phenomenological studies at lower hard scales. We assume that the unpolarized TMD gluon distribution has the following Gaussian form~\cite{Boer:2011kf},
\begin{equation}
f_1^g(x,\bm p_\sT^2) = \frac{f_1^g(x)}{\pi \langle  p_\sT^2 \rangle}\,
\exp\left[-\frac{\bm p_\sT^2}{\langle  p_\sT^2 \rangle}\right]\,,
\label{eq:f1g}
\end{equation}
with a rather large $\langle p_\sT^2\rangle$ = 7 GeV$^2$ to effectively include the broadening effects due to multiple gluon emissions. The 
polarized distribution is chosen to saturate the positivity bound, as in Eq.~(\ref{eq:h1pg}), in order to see the maximal effects allowed. Our results for the transverse momentum distribution of the Higgs plus jet pair, the $\vert\langle \cos 2 \phi\rangle\vert $ and  $\langle \cos 4 \phi\rangle $ asymmetries are shown in Fig.~\ref{fig:gauss}. For the Gaussian model we fix $q_{\sT {\rm max}}^2= K_\perp^2/4$ as explained above.
We find that $\vert \langle\cos 2 \phi\rangle\vert \approx 9\%$ and  $\langle \cos 4 \phi\rangle \approx 0.4\%$ at $K_\perp = 100$ GeV, very similar to the 
values obtained with the previous model, despite the considerable differences. 
At $K_\perp = 30$ GeV, $\vert \langle\cos 2 \phi\rangle\vert \approx 3\%$ and $\langle \cos 4 \phi\rangle \approx 0.02\%$.

As can be seen in the above numerical results, the Gaussian model exhibits a double node in $R_0$, whereas the Gaussian+tail model does not. 
In Ref.\ \cite{Boer:2011kf} it was noted that the TMD convolution involving the weight $w_0^{hh}$ exhibits a double node 
independent of the form of the TMD $h_{1}^{\perp g}$. More explicitly, the following integrals vanish: $\int d^{2}\bm q_{\sT}\, (\bm q_\sT^{2})^\alpha\, \mathcal{C}[w_{0}^{hh}\, h_{1}^{\perp g}\, h_{1}^{\perp g}]=0$ for $\alpha=0$ and $\alpha=1$, which is important to mention is not due to the angular integration. However, this does not imply that the actual distribution $R_0$ exhibits two nodes, because the expression in Eq.~(\ref{eq:R0}) in terms of TMD convolutions only holds for $q_\sT^2 \ll Q^2$ for some hard scale $Q$. In general, $\int_0^{q_{\sT \rm max }^2} \d \bm q_\sT^2 \, (\bm q_\sT^{2})^\alpha \, \mathcal{C}[w_{0}^{hh}\, h_{1}^{\perp g}\, h_{1}^{\perp g}] \neq 0$.
Addition of order $q_\sT^2/Q^2$ terms that are dominant at large $q_\sT^2$ and that can be 
cast into the same convolution form~\cite{Sun:2011iw}, 
allows to extend the integration region to all $q_\sT$. However, this need not lead to a vanishing $ q_\sT^{2}$-weighted integral. In order for linearly polarized gluons to not affect the $q_\sT$-integrated cross section, there should always be one node  at least, but it may well be outside the TMD region, i.e.\ in the region $q_\sT^2 \sim Q^2$. This is why models for $h_{1}^{\perp g}$ can lead to an $R_0$ distribution exhibiting any number of nodes in the TMD region. As the Gaussian model has no significant contributions outside the TMD region, it does have to display two nodes in the TMD region, as we confirm it does. 

\section{Final remarks}
For the measurement of the effects discussed here, the jet transverse momentum resolution is an important factor. As the resolution scale is probably on the multiple GeV level and the effects in the models are largest for $q_\sT$ values in the few GeV region, it may be hard to experimentally probe the region of interest here. We note however that no experimental knowledge is available on the shape of the $h_1^{\perp\ g}$ distribution, hence, there are no constraints available to bound or indicate the typical width of the distribution. In the models we have made specific assumptions ($R= 2$ GeV$^{-1}$ in the Gaussian+tail model and $\langle p_\sT^2\rangle$ = 7 GeV$^2$ in the Gaussian model) which need not correspond to the actual distribution. The latter could be significantly broader than expected from intrinsic transverse momentum effects, due to multiple gluon emissions, just like nonperturbative effects can affect the $Z$-boson production distribution at transverse momentum values well above a few GeV. We therefore caution the reader not to take the $q_\sT$ ranges in the figures too literally. The models are intended to illustrate what kind of features can arise qualitative from linearly polarized gluons. The magnitudes of the asymmetries do give an indication of the maximal effects one might expect.   

Next we comment on possible effects from the color flow in the process. The analysis presented here ignored the effects of initial and final state interactions. In the process $p\, p \to H\, X$, the gauge link structure of $\Phi_g^{[U]\,\mu\nu}(x,\bm p_\sT )$ in Eq.~(\ref{Phig}) is given by two infinite staple-like gauge links that both run to minus lightcone infinity. It is denoted as $\Gamma^{[-,-\dagger]}$ in Ref.~\cite{Bomhof:2007xt}. 
Given that we restrict to $T$-even distributions, this is equal to $\Gamma^{[+,+\dagger]}$. In the process of Higgs plus jet production there is a more complicated gauge link structure, which has not been considered yet for all subprocesses. The gauge link structure of $gq \to H q$ will be the same as that of $g q \to \gamma q$ given in Ref.~\cite{Bomhof:2007xt}, but $g g \to H g$ has no analogue considered before. In any case, we expect that the $h_1^{\perp g [U]}(x, \bm{p}_\sT^2)$ distribution(s) probed in $p\, p \to H\ {\rm jet} X$ are different from the one in $p\, p \to H\ X$. It is not clear at present how large quantitatively the differences between these distributions are in practice, but it should be kept in mind that $T$-even distributions do not require initial or final state interactions to be nonzero, unlike for instance Sivers functions. The gauge link dependence need not be the dominant dependence therefore. As such, there is no reason to assume that they are very large or very small. This simply remains to be seen. Given that there are sufficiently many processes sensitive to the linear gluon polarization, the size of the initial and/or final state interactions can, at least in principle, be extracted from experiment. As a final comment, we point out that there are at present no indications that factorization will be broken in the process $p\, p \to H\,{\rm jet}\, X$ due to color entanglement effects like in $p\, p \to {\rm jet}\, {\rm jet}\, X$ \cite{Rogers:2010dm}. 

In this paper we have considered linear gluon polarization effects in Higgs plus jet production. Compared to similar processes where a color singlet state plus a jet is produced, the case of Higgs production is special. Its large mass allows for an unsuppressed angular independent contribution in the region $q_\sT \ll K_\perp \ll M_H$, cf.\ Eq.~(\ref{eq:R0-2}). This is not the case, for instance, in heavy quarkonium plus jet production, where in addition issues related to color octet contributions (even if suppressed) may lead to complications regarding the factorization \cite{Ma:2014oha}. Other color singlet state plus jet production processes, such as on- or offshell (Drell-Yan \cite{Dooling:2014kia}) photon plus jet production, or $W$ or $Z$ boson plus jet production, are not sensitive to $h_1^{\perp\, g}$, but only to $h_1^{\perp\, q}$. This also applies to Higgs production in association with a photon, $W$ or $Z$ boson. The $H+W$ and $H+Z$ processes have been investigated in great detail already, but without inclusion of polarization effects (of quarks in this case) \cite{Han:1991ia,Hagiwara:1993qt,Chatrchyan:2012qr,Ellis:2013ywa,Godbole:2014cfa,Harlander:2014wda}. We point out that to probe $h_1^{\perp\, q}$ the best processes probably are still the Drell-Yan process as suggested in Ref.~\cite{Boer:1999mm} or photon plus jet production \cite{Boer:2007nd}. Moreover, we note that the process Higgs plus $W$ boson receives a contribution from $h_1^{\perp\, q}$ only proportional to the very small Higgs coupling of the light quarks inside the proton and can therefore rather be viewed as a cross-check process. 

Linear gluon polarization effects in other color singlet pair production processes have been studied, i.e.\ diphoton production in Ref.~\cite{Qiu:2011ai} and $J/\psi$ plus photon production in Ref.~\cite{Dunnen:2014eta}, but as mentioned those studies were done in the Collins-Soper frame, where the restriction $q_\sT \ll K_\perp$ is not imposed or automatically respected. The numerical results obtained in those studies do not show a clear advantage over Higgs plus jet production, giving similar or smaller values for the angular asymmetries. The lower energy and the better transverse momentum resolution could provide a big advantage though.

When it comes to probing the linear polarization of the gluons, Higgs plus jet production has some additional features compared to  inclusive Higgs production. One is the possibility of probing a continuous range of hard scales, allowing, at least in principle, for a study of TMD evolution. This is not possible in Higgs production, where the hard scale is fixed to be $M_H$. Another is that, because the Higgs-jet system can be in various angular momentum states, angular distributions can be probed that are not accessible in inclusive Higgs production where the recoiling jet is not observed. Thanks to this, effects involving only one initial linearly polarized gluon can be probed through the $\cos 2 \phi$ distribution, as in \cite{Boer:2010zf,Qiu:2011ai,Dunnen:2014eta}.

\section*{Acknowledgments}
This work was supported by the European Community under the ``Ideas'' program QWORK (contract 320389). We thank Paolo Gunnellini, Jean-Philippe Lansberg, Hannes Jung and Pierre Van Mechelen for useful comments and information.

\end{document}